\documentclass{JHEP3}
\title{Relativistic  Violation Invariance, Multiverses and Quantum Field Theory}
\author{J. Gamboa\\
   Departamento de F\'{\i}sica, Universidad de Santiago de Chile,
   Casilla 307, Santiago 2, Chile\\
  \email {jgamboa@lauca.usach.cl}}
 \author{M. Loewe\\
    Facultad de F\'{\i}sica, Pontificia Universidad Cat\'olica de Chile,
   Casilla 306, Santiago 22, Chile\\
   \email {mloewe@fis.puc.cl}}
\author{F. M\'endez
   \\
   INFN, Laboratori Nazionali del Gran Sasso, SS, 17bis, 67010 Asergi (L'Aquila),  Italy
   \email{ fernando.mendez@lngs.infn.it}}
\abstract{The possibility of interaction among multiverses is studied assuming that in the first instants of the  big-bang,  many disjoint regions were created producing many independent universes (multiverses). Many of these mini-universes were unstable and they decayed, but other remained as topological remnant (like domain walls or baby universes) or possibly as mini-black-holes. In this paper, we study the quantum statistical mechanics of multiverses assuming that in the first instants of the big-bang,  the relativistic symmetry was only an approximate symmetry and the interaction among multiverses was produced by non-local communication. As a warm-up, in the first part of this paper we  study the statistical quantum mechanics of generally covariant systems (particles, strings and membranes) living on noncommutative spaces.
   In the second part, the non-local communication is implemented by noncommutativity in the fields space and the possible physical implications in cosmology are considered.   As the Lorentz symmetry is broken,  technically the problem is solved assuming  a privileged  reference frame containing the multiverses, {\it i.e.} a kind of ideal quantum gas in a reservoir. If the reservoir is very large, then one can consider a uniform multiverses distribution and approximate {\em each multiverse as tensionless p-brane}. The breaking of the relativistic symmetry induces on each multiverse a tiny harmonic interaction. The oscillation frequency for each multiverse is proportional to $1/B$, where $B$ is the  noncommutativity parameter. We argue that $B$ can identified as the primordial magnetic field,  {\it i.e.}  $\sim 10^{-16}\, \mbox{GeV}^2$. In this model of multiverses, each multiverse interacts with other neighbour multiverses in a similar way as atoms do in the Einstein model for the specific heat of a solid. In this case, the analogous of the phonon is played by  quantums with energy equal to $B$. Each neighbour multiverse should have a  pulsation frequency  $\Omega \sim 10^{-63} \,s^{-1}$.  This tiny frequency could suggest that the relativistics invariance --from the cosmological point of view--  is almost exact and the multiverses could be not detected using the presently astronomical observations.}

\begin{document}

   \section{Introduction}
   \subsection{Motivation and Physical Implications}

   Presently there is a strong observational evidence in favor of the big-bang theory \cite{pee}.
   However, there still remain many open problems that could reinforce
   this evidence if we are able to find simple explanations for baryogenesis, leptogenesis, primordial magnetic fields and so on.

   Several years ago, people suggested, as a consequence of the big-bang theory, that objects like monopoles, cosmic strings and domain walls could remain in the present universe as topological remnants \cite{review1}. In this context, monopoles and cosmic strings were studied extensively in the eighties and several possible observational procedures were proposed \cite{review2}.

   However, domain walls are still not well understood because these objects produce causally disconnected spacetime regions and the physical possibility of detecting these domain walls, using the presently relativistic standard cosmology, is impossible.

   A variant of this problem and -apparently an unrelated problem- is the multiverse picture advocated  by Linde \cite{linde1}  where the possibility of new many universes is allowed.

   In some sense, however, these two perspectives, namely domain walls and multiverses, could be just two different semantic aspects of the same problem if the relativistic invariance in the first instants of the universe is assumed to be exactly. However, if this fact is true, then  the microcausality principle would make impossible to detect signals coming from different causally disconnected regions.

   One should emphasize, however, that the concepts of domain wall, multiverses, baby universes and so on, could be,  indeed, a physical manifestation of the first instants of the universe. In the chaotic inflation scenario,  many big-bangs could have taken place during the first period of the creation  and, as consequence, disconnected spacetime regions should have emerged. So,in a semantic sense, many independent universes were created.

   In this paper, the concept of \lq \lq many universes" (or multiverse) is understood in this sense.

   An important question is if multiverses are only a theoretical construction or if they could be physically detected?

  In this paper, we propose the possibility of detecting interactions among  multiverses  assuming that, during the first instants of the universe,  the relativistic symmetry was only approximate.  As a consequence, causally disconnected  regions of spacetime exchanged information after this tiny relativistic invariance violation took place.

  In this approach, there are many points that should be answered and, of course, one of them is,  how this relativistic invariance violation occurs? If in the first instants of the universe the relativistic invariance was not an exact symmetry then, which are the relics in the present universe?.

  Recently, the possibility of having a violation of the Lorentz invariance has been discussed extensively in different particle processes as, for example, adding small not Lorentz invariance  terms to the lagrangian \cite{kos}, modifying the dispersion relations \cite{alot} \footnote{The earlier papers on double special relativity appeared in \cite{ame1}.} or using general arguments coming from noncommutative geometry.

  In our opinion,  the violation of the relativistic symmetry can be a real possibility because -after all-  the presently relativistic quantum field theories are only effective descriptions of nature. They could be valid only up to energies $\theta \sim 1/M_p$, where  the cutoff $M_p$ is the Planck mass.

  In a couple of previous papers \cite{ccgm}, we have discussed a different alternative possibility, namely a noncommutative realization in the field space where violation of the  Lorentz invariance appears because the microcausality principle is no longer valid.

  This assumption allows to find  a geometric explanation for purely phenomenological dispersion relations and it also provides a natural explanation for the different non-relativistics terms that one could add to the lagrangean.

  \subsection{Technicalities and quantum field theory}

   The idea of noncommutative fields is interesting because it might imply important consequences for our explanations of the physical world. Although noncommutativity in the field space induces a violation of the microcausality principle, it also provides an explicit mechanism for non-local communication at the quantum field theory level.

As is well known, in quantum field theory \cite{weinberg} the cluster decomposition principle  states that if two events are sufficiently separated, then the hamiltonian density satisfies
\[
\left[{\cal H}({\bf x}), {\cal H}({\bf y}) \right] = 0,
\]
if the separation $(x-y)^2 <0 $ is spacelike.

The hamiltonian density ${\cal H}$ is a local functional of the fields $\phi_1, \phi_2, ...$ and, therefore, the microcausality principle, namely
\begin{equation}
   \left[ \phi_i ({\bf x},t), \phi_j ({\bf x}^{'},t)\right] =0, \label{3}
\end{equation}
is simply a consequence of  the cluster decomposition principle.

From the mathematical point of view, the cluster property assures causality and Lorentz invariance. Otherwise (\ref{3}) they cannot be realized \cite{weinberg} .

If one takes a different point of view, namely, relaxing the cluster property, then an experiment in a given point ${\bf x}$ could affect -in principle in a severe way- another experiment in another point ${\bf y}$.

This last fact is known as non-local communication. Let us
consider physical systems such that they can interact via a
non-local communication mechanism. Technically this interaction is
implemented assuming {\it mutatis mutandis} that (\ref{3}) can be
deformed as
\begin{equation}
   \left[ \phi_i ({\bf x}), \phi_j ({\bf x}^{'})\right] = i \theta_{ij} ~
   \delta ( {\bf x} -{\bf x}^{'}),
   \label{4}
   \end{equation}
   where $\theta_{ij}$ is a constant antisymmetric matrix that measures the causality violation.

   In analogy with many body theory, and by simplicity, we can assume that the non-local communication is paired. So we can replace  the general antisymmetric matrix $\theta_{ij}$ by $\epsilon_{ij} \theta$. In this description, if the energy scale is , say,  $\Lambda$, then the dimensionless quantity $\Lambda \theta<<1$ \footnote{we are assuming, for simplicity, scalar fields in a four-dimensional spacetime.} is very small, being a measure of a tiny causality violation.

   The deformation of the canonical algebra of fields yields to very interesting phenomenological consequences such as: a) non-trivial dispersion relations \cite{ccgm}  that could explain cosmic ray physics and the violation of the GZK cutoff \cite{gzk}, b) a possible explanation of the matter-antimatter asymmetry \cite{rubi}, c) a new approach to phenomenological relics of quantum gravity \cite{alot}.

   This non-local communication induces a violation of relativistic invariance and, as it was discussed above, this point of view could be useful for discussing several open cosmological problems such as:

   a) Communication among causally disconnected spacetime regions, {\it e.g.} domain walls

   b) Interaction between multiverses. This point has been extensively discussed by Linde where he finds a global  interaction between two universes, retaining relativistic invariance. In our case, we will find a non-local interaction that, eventually, could produce a very small gravitational radiation.

   c) Using dimensional arguments we could also discuss another elusive problem, namely, the primordial magnetic field.

   The equation (\ref{4}) is the definition of the {\it noncommutative fields} as an extension of
   \[
   \left[ x,y\right] = i~ \theta,
   \]
   in noncommutative field theory.

   The equation (\ref{4}) means that two regions that are causally disconnected, could exchange information if we allow a small noncommutativity  in the phase space of fields.  This fact means that, eventually,  non- interacting systems could in fact interact due to the noncommutative structure of the phase space of field. Thus, noncommutativity provides a very natural way  for introducing interactions.

\subsection{Physical Discussion}

Although this last fact might be a problem in a relativistic theory, in our case  the Lorentz invariance is explicitly broken once (\ref{4}) is assumed and, therefore, it is interesting to consider physical realizations of such possibility.

A realization of this last fact occurs -as was discussed above- in cosmology and in this subsection we will motivate the  problem explicitly.

Indeed, since the gravitational field is given by the metric,  the
microcausality principle would imply
\begin{equation}
\left[ g_{ij}({\bf x}), g_{kl}({\bf x}^{'})\right] = 0. \label{ww}
\end{equation}

Thus, one concludes that the conventional gravity theory cannot have non-local communication unless one breaks the explicitly the relativistic invariance (\ref{ww}). Possibles interactions among 
causally disconnected regions of the spacetime by relativistic invariance is, of course, forbidden. 

If we admit that relativistic invariance is broken in the sense of (\ref{4}), then our universe is only one of the many possible universes contained in a sort of reservoir, {\it i.e.} a gas of universes. In this reservoir one can define an evolution  parameter $s$ which may coincide with the conventional time and, therefore, would allow  to define an evolution operator as in quantum mechanics.

   From this point of view, let us assume that the universe $i$ is described by a metric $g$ and has a field  $\Psi_i(g)$, then, the condition for non-local communication  among universes is
   \begin{equation}
   \left[ \Psi_i (g), \Psi_j (g^{'}) \right] = i \theta_{ij} \delta (g,g^{'}),
   \label{5}
   \end{equation}
   with $i,j= 1, 2, 3, \dots$.

   The possibility $\theta_{ij}\neq 0$ could give information about the existence of other universes  and it would provide an evidence for  the violation of the causality principle at very high energies.

   The purpose of this paper is, firstly,  to construct noncommutative versions of generally covariant systems and, secondly, to elaborate the approach sketched above for cosmology, exploring the consequences of a weak violation of the causality principle.

   The paper is organized as follow: in section II we consider relativistic particles  in a noncommutative space analizing several quantum statistical mechanics considerations. In section III, the non-commutative $p$-dimensional membrane in the strong coupling limit and the quantum statistical mechanics of these systems is studied. In section IV, we  propose a non-local communication mechanism for particles in quantum mechanics. In section V, we apply our previous considerations to cosmology and, finally in section VI we present our conclusions and outlook.

   \section{Noncommutative relativistic quantum mechanics} \label{sec:NC1}

   In this section we will construct noncommutative versions of generally covariant systems. We will start considering, firstly, the relativistic particle on a $D$-dimensional spacetime and later-- in the next section -- we will extend our results to tensionless strings and membranes.

   \subsection{Relativistic free particle and the proper-time gauge}

   There are many approaches to discuss relativistic quantum mechanics of a free particle. One of them is the so called proper-time method, which was used   in the early 50th in connection with quantum electrodynamics \cite{nambu}. The idea is to consider a particle in a $D+1$- dimensional Euclidean  spacetime.

   The diffusion equation for such system is
   \begin{equation}
   -\frac{1}{2} \Box \varphi (x, s) =  \frac{\partial \varphi}{\partial s},
   \label{6}
   \end{equation}
   where $\Box$ is the $D$-dimensional Laplacian.

   Then, using the ansatz
   \begin{equation}
   \varphi (x,s) = e^{-\frac{m^2}{2} s} \phi (x), \label{7}
   \end{equation}
one finds that $\phi(x)$ satisfies the Klein-Gordon equation if $m$ is the mass of the particle.

In this approach, the propagation amplitude is  given by the Laplace transform
   \begin{equation}
   G[x,x^{'}; m^2] = \int_0^\infty d s e^{-s\frac{m^2}{2} } \, G[x,x^{'}; s],
   \label{8}
   \end{equation}
   where
   \begin{eqnarray}
   G[x,x^{'}; s] &=& \int {\cal D} x \, e^{- \int_0^1 d \tau \frac{{\dot
   x}^2}{2 s}}, \label{9} \nonumber
   \\
   &=& s^{-D/2} \, e^{- \frac{(\Delta x)^2}{2 s}}.\label{99}
   \end{eqnarray}

   From this one obtains the partition function for a gas of $N$ free relativistic particles \footnote{In this paper we will ignore  the Gibbs factor $1/N!$. The reader should note also that we are assuming the Maxwell-Boltzmann statistics.}
   \begin{equation}
   Z_s=\left(\mbox{Tr}\left[e^{-\frac{m^2}{2} s} G[x,x';s]\right]\right)^N,
   \end{equation}
   or equivalently
   \begin{equation}
   \ln Z = N \left[ - \frac{m^2}{2} s- \frac{D}{2} \ln s   + \ln {\cal V} \right],
   \label{10}
   \end{equation}
   where ${\cal V}=V \times \mbox{const.}$ is the $D$-dimensional spacetime, $V$ is the
   $D-1$-dimensional ordinary spatial volume and $s$ plays the role of $\beta =1/kT$.

   \subsection{The relativistic particle in a noncommutative space}

   Equation (\ref{6}) suggests a simple way to extent the problem to a gas of relativistic particles on a noncommutative space.

   Indeed,  from (\ref{6}) we see that the Hamiltonian for a relativistic particle is
   \begin{equation}
   {\hat H} =\frac{1}{2} p^2_\mu.  \label{ham}
   \end{equation}

   Once (\ref{ham}) is given, noncommutativity is implemented through the deformed algebra
   \begin{eqnarray}
   \left[ x_\mu, x_\nu\right] &=& i \theta_{\mu \nu},
   \,\,\,\,\,\,\,\,\,\,\, \left[ p_\mu, p_\nu\right] = i B_{\mu \nu}, \label{11}
   \\
   \left[ x_\mu, p_\nu\right] &=& i \delta_{\mu \nu}, \label{12}
   \end{eqnarray}
   where $\theta_{\mu \nu}$ and  $B_{\mu \nu}$ are the deformation parameters in the phase space.

   By convenience  we choose the gauge
   \begin{eqnarray}
   \theta_{i0} &=& 0\, \, \, \,\, \, \, \,  \theta_{ij} = \epsilon_{ij}
   \theta, \label{13}
   \\
   B_{i0} &=& 0, \, \, \, \,\, \, \, \, B_{ij} = \epsilon_{ij}  B.
   \label{14}
   \end{eqnarray}

   Therefore, the equation of motion for this particle is
   \begin{eqnarray}
   {\dot x}_\mu &=& p_\mu, \nonumber
   \\
   {\dot p}_i &=&  \epsilon_{ij} B p_j.  \label{15}
   \end{eqnarray}

   These equations can be integrated directly by using (\ref{13}) and (\ref{14}). Indeed, one of the equations is trivial, namely, the energy conservation condition ($\dot{p}_0=0$). Note that the symmetric gauge we have chosen, implies that non-commutativity is realized only for the first two momenta and coordinates components. The other components are treated as usual. In principle, we could extend this hypotesis taking also other pairs of momenta and coordinates components, but this is not essential for our discussion.

Keeping this in mind, the remaining equations have the solution
   \begin{eqnarray}
   p_1 &=& \frac{1}{2} \left( \alpha~ e^{-i B t} + \alpha^\dag ~ e^{ i B t}
   \right), \nonumber
   \\
   p_2 &=& \frac{1}{2i} \left( \alpha ~e^{- i B t} - \alpha^\dag~ e^{ i B
   t} \right), \label{16}
   \end{eqnarray}
   where  $\alpha$'s are constant operators.

   The coordinates $x_ {1,2}$ are obtained in a similar way using (\ref{15}), {\it i.e.}
   \begin{eqnarray}
   x_1 &=& \frac{1}{2 i B} \left(\alpha^\dag~ e^{ i B t} -\alpha~ e^{- i B t}  \right) + x_{01}, \nonumber
   \\
   x_2 &=& \frac{1}{2  B} \left( \alpha~ e^{- i B t} + \alpha^\dag~ e^{ i B t} \right)+ x_{02}. \label{17}
   \end{eqnarray}

   From the commutation relation of $p$'s, we see that it is possible to define operators $a$ and $a^{\dagger}$ satisfying the algebra
   \begin{eqnarray}
   \left[a, a \right] &=& 0=\left[a^\dag, a^\dag \right], \nonumber
   \\
   \left[a, a^\dag \right] &=& 1 \label{18},
   \end{eqnarray}
   where
   \[
   \alpha \rightarrow \sqrt{B} a,\,\,\,\,\,\,\,\,\,\,\,\,\,\,\,\,\,\, \alpha^{\dag}
   \rightarrow \sqrt{B}a^{\dag}.
   \]

   The equations of motion --as a second order equation system-- are
   \[
   \ddot{x}_\mu=B_{\mu\nu}\dot{x}_\nu,
   \]
   which can be solved by the Ansatz $x_\mu = a_\mu \,e^{i\omega s}$.

   The last equation is
   \[
   (i\omega\delta_{\mu\nu} -B_{\mu\nu})a_\nu=0.
   \]

   Therefore, the dispersion relation for this system is
   \begin{equation}
   \omega_{\pm}=\left\{
   \begin{array}{c}
   \pm B \\
   0
   \end{array}
   \right. , \label{dis}
   \end{equation}
   and --since one of the eigenvalues vanishes-- the Hamiltonian spectrum is degenerated

   Thus, the  hamiltonian for a relativistic particle living on a noncommutative space is
   \begin{equation}
   H=\frac{B}{2}\left( a^\dag a +\frac{1}{2}\right)
   +\frac{1}2\sum_{n=1}^{D-3}(p_\mu^2)_n.
   \label{23}
   \end{equation}

   Finally, the statistical mechanics for a gas of $N$ relativistic particles on a noncommutative space, in the symmetric gauge, is obtained from the partition function
   \begin{eqnarray}
   Z_s&=&\left( s^{-\frac{D-3}{2}}e^{-\frac{m^2}{2}s} \sum_{n=0}^{\infty}
   {\cal G}_0 e^{-s\frac{B}{2}(n+\frac{
   1}{2})} \right)^N,\nonumber
   \\
   &=&\left[\frac{{\cal G}_0   e^{-\frac{m^2}{2}s}
   s^{-\frac{D-3}{2}}}{\sinh{(\frac{B}{2} s)}} \right]^N,
   \label{25}
   \end{eqnarray}
   where ${\cal G}_0$ is the degeneracy factor due to the zero eigenvalue of the Hamiltonian \footnote{Although this factor can be computed by using a regularization prescription, here this factor is absorbed as a normalization constant.}.

   The thermodynamic properties of this system can be computed directly from (\ref{25}).

   \section{The strong coupling regime for membranes in noncommutative spaces} \label{sec:NC2}

   In this section we will discuss the extension  of the previous problem to membranes moving on a noncommutative space in the strong coupling regime.

   A relativistic membrane is a $p$-dimensional object embedded on a $D$-dimensional flat spacetime
   and described by the lagrangean density
   \[
   {\cal L} = \frac{1}{2} \sqrt{g^{(p+1)}} \left[g_{\alpha \beta} G^{\mu \nu} \partial^{\alpha} x_\mu  \partial^{\beta} x_\nu -(p-1) \right],
   \]
   where $g^{(p+1)}_{\alpha \beta}$ ($\alpha, \beta= 0, 1, 2, ...p$) is metric tensor on the world-volume and
   $G^{\mu \nu}$ is the metric tensor where the $p$-brane is embedded with $\mu, \nu = 0,1,2, ...,D$.

   The hamiltonian analysis yields to the following constraints
   \begin{eqnarray}
   H_{\perp}&=&\frac{1}{2}(p^2 + T^2 g^{(p)}),\label{26}
   \\
   H_i&=&p_\mu\partial _i x^\mu, \label{27}
   \end{eqnarray}
where $g^{(p)}$ is the spatial metric determinant and $T$ is the superficial tension.

   The strong coupling regime corresponds to $T\rightarrow 0$ and, in this limit the constraints are
   \begin{eqnarray}
      H_{\perp}&=&\frac{1}{2} p^2 ,\label{2666}
      \\
      H_i&=&p_\mu\partial _i x^\mu, \label{2777}
   \end{eqnarray}
   and the membrane becomes an infinite set of free massless relativistic particles moving perpendicularly  to the $p$-dimensional surface.

   In the special case of the tensionless string ($p=1$), each point of the string is associated with  a massless relativistic particle and, as a consequence, all the points of the string are causally disconnected.

   In this tensionless string approach the field $x^\mu(\sigma ,\tau)$ is replaced by $x^\mu_i (\tau)$, where $i=1, 2, ...,$ is  an infinite countable set labeling each point of the tensionless string.

   Using this philosophy, we will start constructing tensionless strings.

   \subsection{Tensionless strings from particles}

   Let us start by noticing that a tensionless string \cite{gamboa} is made up of infinite massless relativistic particles causally disconnected and, therefore, instead of (\ref{6}) one have
   \begin{eqnarray}
   -\frac{1}{2}\Box \varphi_1 (x, s_1)&=& \frac{\partial
   \varphi_1}{\partial s_1},\nonumber
   \\
   -\frac{1}{2}\Box \varphi_2 (x, s_2) &=&  \frac{\partial
   \varphi_2}{\partial s_2},\nonumber
   \\
   &\vdots& \nonumber
   \\
   -\frac{1}{2}\Box \varphi_k (x, s_k) &=&  \frac{\partial
   \varphi_k}{\partial s_k}.
   \end{eqnarray}

   These equations can be solved  by generalizing the Ansatz (\ref{7}), {\it i.e}
   \begin{equation}
   \varphi(x_1,\dots,x_k,\dots;s_1\dots,s_k,\dots)=\prod_{i=1}^{\infty}~e^{-\frac{m^2}{2}s_i}\phi(x_i),
   \end{equation}
where $m^2$ is an infrared regulator that will vanish at the end of the calculation.

  The limit  of  an infinite number of particles  is delicated but here --formally-- one can take this limit, simply, assuming that in the continuous limit one can replace the set $\{ i\}$ by an integral in $\sigma$ and, as a consequence,   the propagation amplitude can be written as:
   \begin{eqnarray}
   &G&[x(\sigma),x'(\sigma)]=\nonumber
   \\
   &=& \int_0^{\infty} {\cal D} s(\sigma)~e^{-\frac{m^2}{2}\int  d\sigma
   s(\sigma)}G[x(\sigma),x'(\sigma);s(\sigma)],  \label{29}
   \end{eqnarray}
   where $G[x(\sigma),x'(\sigma);s(\sigma)]$ is given by
   \begin{equation}
   G[x(\sigma),x'(\sigma);s(\sigma)]=s^{-D/2}(\sigma)  e^{-\int
   d\sigma \frac{[\Delta x(\sigma)]^2}{2s(\sigma)}}. \label{299}
   \end{equation}

The formula (\ref{29}) generalizes the proper-time method to the tensionless string case. Probably this approach to string theory was first used by Eguchi in \cite{gamboa}.

   Using (\ref{29}) and (\ref{299}), the partition function of an $N$ tensionless string gas is
   \begin{eqnarray}
   Z[s(\sigma)]&=&\left[ \int {\cal D} x(\sigma)\,  G[x(\sigma),x(\sigma);s(\sigma)]\right]^N\nonumber
   \\
   &=& \left(s^{-D/2}~e^{-\int d\sigma
   \frac{m^2}{2}s(\sigma)}\right)^N.
   \label{32}
   \end{eqnarray}

   This partition function reproduces correctly the results for the thermodynamics  of a tensionless string gas \cite{string}.

   Indeed, from (\ref{32}), the Helmholtz free energy is
   \[
   F[s]=\frac{N}{s(\sigma)}\left[ \frac{D}{2} \ln(s(\sigma)) +\frac{m^2}{2}\int
   d\sigma s(\sigma) +\ln({\cal V})
   \right].
   \]

   As  $1/s$ is the temperature, then from the limit $m^2\rightarrow 0 $ we see that $F/T \sim \ln(T)$, again in agreement with other null string calculations \cite{string,atick}.

   From the last equation one obtain that
   \begin{equation}
   P[s(\sigma)]{ V}=\frac{N}{s(\sigma)},
   \end{equation}
   is the state equation for an ideal tensionless string gas.

   \subsection{Tensionless membranes from tensionless strings} \label{sec:NC3}

  In order to construct tensionless membranes, we begin by considering a membrane as an infinite collection of tensionless strings.
  Thus, if the membrane is a p-dimensional object, with local coordinates $(\sigma_1,\dots,\sigma_p)$,
  then the propagation amplitude, formally, corresponds to  (\ref{29}), with the substitution
   \[
   \sigma\rightarrow (\sigma_1,\dots,\sigma_p).
   \]

   Therefore, the partition function for a gas of $N$ tensionless membranes is
   \begin{eqnarray}
   Z[s(\sigma)]= \left[\lim_{n \to \infty}\left([s(\sigma)]^{-D/2}~
   e^{- \frac{m^2}{2}\int d^p \sigma s(\sigma)}{\cal V}
   \right)^n\right]^N, \nonumber \label{34}
   \\
   &&
   \end{eqnarray}
   where $n$ is the number of tensionless strings.

   One should note here that the expression
   \[
   \left([s(\sigma)]^{-D/2}~e^{- \frac{m^2}{2}\int d^p \sigma s(\sigma)}\right)^n,
   \]
   formally emphasizes that a tensionless $p$-branes is made-up of $n$ tensionless strings.

   However, this last expression was computed in (\ref{299}) and in our case is
   \[
   \prod_{i=1}^p[s(\sigma_i)]^{-D/2} \, e^{- \frac{m_i^2}{2}\int d \sigma_i s(\sigma_i)},
   \]
then, the total partition function for an ideal gas of $N$ tensionless p-branes is given by
\[
Z =  \prod_{i=1}^p \left([s(\sigma_i)]^{-D/2}~ e^{-\frac{1}{2}m_i^2 \int d\sigma_i s(\sigma_i)}\right)^N.
\]

In order to compute the state equation we proceed as follow:
firstly one chooses $s(\sigma_1)=s(\sigma_2) ...=s(\sigma)$ and
one put also $m_1=m_2= ...=m$, then
   \begin{equation}
   P[s(\sigma)]~{ V} = \frac{N}{s(\sigma)}.
   \end{equation}

   The Helmholtz free energy,  compared to the tensionless string case, has a different behavior. Indeed, the Helmholtz free energy becomes
   \[
      F[s]=\frac{pN}{s(\sigma)}\left[ \frac{D}{2} \ln(p\,s(\sigma)) +\frac{m^2}{2}\int
      d\sigma s(\sigma) +\ln({\cal V}).
      \right].
   \]
   and  for $s\rightarrow \infty$, one has that the quantity $sF \sim \frac{D}{2}  \ln [p\, s]$ is similar to the string case but, in this case $p$ could smooth out  the behaviour of $sF$.

   \subsection{Including noncommutativity in Tensionless p-branes}

Using the previous results,  we can generalize our arguments in order to include noncommutativity in tensionless p-branes. In order to do that, one start considering a tensionless p-brane described by the {\bf field} $x_i^\mu (\tau)$ with $i$ labeling  the dependence in $(\sigma_1,\sigma_2, ..\sigma_p)$.  This field transforms as a scalar on the world-volume but as a vector in the space where the p-brane is embedded.

Let us suppose that the components --we say $x_i^{{D-1}}$  and $x_i^D$-- do not commute,
then -in such case- the Green function can be written as
\begin{eqnarray}
&&
G[x(\sigma),x'(\sigma);s(\sigma)] = \nonumber
\\
&=& \int_0^\infty  ds\, e^{-\frac{m^2}{2} s}\prod_{k=0}^{D-3} \left[ \int {\cal D} x_i^k \,
e^{-\int_0^1d \tau \,\frac{1}{2s} ({\dot x_i^k})^2 }\right]
 \nonumber
\\
&\times& \int {\cal D} {x_i}^{(D-2)} {\cal D} {x_i}^{(D-1)} \, e^{-\int_0^1 d \tau  \frac{1}{2s} \left(({\dot x_i^{(D-2)}})^2 +
({\dot x_i^{(D-1)}})^2 \right)}. \nonumber
\\
\label{gr1}
\end{eqnarray}

The  integral in the second line in the RHS, corresponds formally
to a non-relativistic particle with mass ($s^{-1}$) moving in
plane in the pressence of a constant perpendicular magnetic field
$B$. In the first line in the RHS,  however,  the integral
formally correspond to the Green function for a set of $p$ free
relativistic particles moving in $(D-3)$-dimensional spacetime.

Thus, the calculation of these integral is straightforward. Indeed, 
\begin{eqnarray}
\int_0^\infty ds [s(\sigma)]^{-\frac{D-3}{2}}&~&s(\sigma) \,e^{ -\frac{1}{2s}(\Delta x^k_i)^2 - \frac{p}{2}m^2 \int d\sigma s(\sigma)} \nonumber
\\
&\times& \mbox{H. O.}, \nonumber
\end{eqnarray}
where $\mbox{H. O}$ means the harmonic oscillator calculation for the two-dimensional relativistic Landau problem.

Thus, the partition function for this gas of $N$-tensionless $p$ branes
   \begin{eqnarray}
   Z[s (\sigma)]&=&\mbox{Tr}\left[G[x(\sigma),x'(\sigma);s(\sigma)]
   \right]\nonumber
   \\
   &=&\biggl( [s(\sigma)]^{-\frac{D-3}{2}}e^{-\frac{pm^2}{2}\int d\sigma
   s(\sigma) }  \sum _{n=0}^{\infty}{\cal G}_0 e^{-
   \frac{B}2(n+\frac{1}{2}\int d^p\sigma s(\sigma))} \biggr)^{N} \nonumber
   \\
   &=&\left[ \frac{{\cal
   G}_0[s(\sigma)]^{-\frac{D-3}{2}}}{\sinh\left(\frac{p\,B}{2}\int d\sigma s
   (\sigma)\right)}\right]^{N}. \label{la}
   \end{eqnarray}

   Therefore, if we assume pairing interaction, then noncommutativity induces a motion for a tensionless $p$-branes confined via a harmonic potential oscillator.

   \section{Interactions via noncommutativity in the phase space} \label{sec:NC4}

   In the previous section we argued how to construct noncommutative  extended objects.
   In this section we would like to give an insight in a different physical context and
   to investigate the possibility of non-local communication.

   Physically speaking, this is a delicate point in the context of a relativistic quantum
   field theory because --as was discussed in the introduction-- the cluster property
   prohibits
   non-local communication and, as a consequence, the microcausality principle is no longer valid.

   From the non-relativistic point of view, apparently there
   is no problem with non-local communication  \cite{peres}. Indeed,  let us suppose two non-relativistic particles in one dimension, labeled by coordinates $x_1$ and $y_1$ and canonical momenta $p_1$ and $p_2$ respectively. Note thate the index refers now to the particles involved.

   The Hamiltonian for this system is
   \begin{equation}
   H=\frac{1}{2}p_1^2 + \frac{1}{2}p_2^2.
   \label{38}
   \end{equation}

   Although naively the particles in (\ref{38}) are free, they can interact if we posit the commutator
   \begin{equation}
   [p_1,p_2]=iB,
   \label{39}
   \end{equation}
   where  $B$ measures the strength of this interaction which can play --or not--the role of a magnetic field.

   Therefore, if (\ref{39}) is fulfilled, then the two-particle system (\ref{38}) is equivalent to an effective {\bf one particle}
   living in a two-dimensional noncommutative space. The exact equivalence between this system and
   the Landau problem is a subtle point because by considering only a noncommutative phase
   space with noncommutative parameters $\theta$ and $B$, one can show
   that noncommutative quantum mechanics and the Landau problem coincide if the relation $\theta = 1/B$ is fulfilled, i.e.
   if we have just the magnetic lenght                                                                                                                                  \cite{gamboa1}. From this example, one extract as conclusion that the equivalence between a physical system such as the Landau problem and noncommutative quantum mechanics only occurs for the critical point $\theta B =1$, but for differents values of $\theta B$, noncommutative quantum mechanics decribes a physics completely different from the Landau problem.  What kind of physics?, presently we do not know the answer to this question.

   The above example can be generalized for more particles; for instance, let us consider two free particles moving in a commutative plane.

   The Hamiltonian  is
   \begin{equation}
   H=\frac{1}{2}(p_{1x}^2+p_{1y}^2)+\frac{1}{2}(p_{2x}^2+p_{2y}^2).
   \end{equation}

   Then, let us assume that the interaction is given by \footnote{Of course this a simplification because we are assuming that the noncommutative parameters are the same.}
    \begin{equation}
    [p_{1x},p_{2x}]=iB,\,\,\,\,\,\,\,\,\,\,\,\,\,\,\,[p_{1y},p_{2y}]=iB,
   \label{40}
    \end{equation}
    then, as in the previous case,  the Hamiltonian is
   \begin{equation}
   H=\frac{1}{2}(p_{1x}^2+ p_{2x}^2)+\frac{1}{2}(p_{1y}^2+ p_{2y}^2),
   \label{41}
   \end{equation}
   in other words, the previous system can be understood as an effective {\bf two particles} system living on a noncommutative plane.

   Thus, the commutator (\ref{40}) and the hamiltonian (\ref{41}) describe a couple of particles living on a plane and interacting formally with a magnetic field perpendicular to the plane.

   In the general case for $N$ particles moving on a $D$ dimensional  commutative space, the generalization is straightforward.

   Indeed,  the Hamiltonian is
   \begin{equation}
   H =  \frac{1}{2} (p^2_{1x} +p^2_{1y}+ ...) + \frac{1}{2} (p^2_{2x} +p^2_{2y}+ ...) + ... ,  \label{l6}
   \end{equation}
    then the interaction can be written as
   \begin{equation}
   [ p_i^a, p_j^b] = i \delta_{ij} \epsilon^{ab} B, \label{l7}
   \end{equation}
   where $a,b$ run on  $1, ...,D$ labelling the different species of particles and the indices $i,j, ...$ select the vectorial component of
   ${\bf x}$  \footnote{The component of the antisymmetric density tensor $\epsilon^{ab}$ are  defined as $+1$ if $a>b$.}.

   If we rewrite the Hamiltonian as
   \begin{equation}
   H =  \frac{1}{2} (p^2_{1x} +p^2_{1x}+ ...) + \frac{1}{2} (p^2_{1y} +p^2_{1y}+ ...) + ... ,  \label{l8}
   \end{equation}
   then, (\ref{l8}) is equivalent to {\bf ${\bf D}$ particles} moving on a $N$-dimensional  noncommutative space.

   In the critical point, this generalized system corresponds to the quantum Hall effect as has been proposed using a different argument by \cite{poly2}.

   Thus,  in a general context, one could conclude that if two particles interact via non-local communication, then the phase space is necessarily noncommutative.

   \section{Cosmological implications}\label{sec:NC4}

   The goal of this section is to use the previous results in order to understand several cosmological issues mentioned in the introduction.

   The first issue to be discussed is how to detect --if this is possible-- causally disconnected spacetime regions. Let us suppose two spacetime regions --say ${\cal S}$ and ${\cal S}^{'}$-- where two experiments take place.  As we know, the cluster decomposition principle states that if the physics for these experiments is described by the hamiltonian densities
   ${\cal H}$ and ${\cal H}^{'}$ defined on ${\cal S}$ and ${\cal S}^{'}$, respectively, then the commutator
   \[
   \left[ {\cal H}, {\cal H}^{'}\right] =0,
   \]
implies that the regions ${\cal S}$ and ${\cal S}^{'}$ are independent and causally disconnected.

In the context of standard relativistic quantum field theory, this property assures relativistic invariance and, in the special case where the vector $x-y$ is spacelike, the microcausality principle emerges as a consequence of this cluster property.

Following standard arguments \cite{linde1,coleman}, in the first stages of the universes probably  many big-bang processes took place and, as a consequence, many baby universes or multiverses were created. These multiverses produced many causally disconnected regions in spacetime. Thus, an important  question is, are really these regions causally disconnected or some tiny interactions could had been possible?.

Presently, we do not have a definitive answer for this question, but there are some phenomenological clues supporting this possibility.

Indeed, in the present epoch of our universe, the relativistic
invariance seems be an exact symmetry but, in the initial stages
--at Planckian energies-- a tiny violation could have been
possible. If we consider the physics of cosmic rays,  for example,
this tiny violation of the relativistic symmetry could explain the
modifications to the GZK bound observed in several recent events
\cite{swain}.

Many people have tried to explain these events  using nonconventional physics \cite{kos,alot,ccgm}.

Another problem  --insoluble problem using the standard
cosmological model-- is the existence of strong magnetic fields in
galaxies, (this problem, probably is related to the
baryon-antibaryon asymmetry) where, apparently, explanations based
on considerations beyond  the relativistic physics are also
necessary.

Thus, assuming that a tiny violation of the cluster principle is
allowed, if the multiverses exchange information, a  tiny
non-local communication among multiverses is also allowed.

The cosmological principle implies that --at very large scales--
the universe is homogeneous and isotropic and, as a consequence, one could think that
each  multiverse is a bubble that --independently-- obeys the cosmological principle, in
other words, one can assume that each constituent universe is a tensionless p-brane.

  Relaxing the cluster decomposition principle,  the multiverses  interact via non-local communication, {\it i.e.}
  noncommutatively  and, as consequence, the interaction among
  multiverses will be possible only if relativistic  invariance is explicitly violated.

   In order to produce interaction among different tensionless $p$-branes, one breaks relativistic invariance assuming nontrivial commutators --like (\ref{l7})--  for the infinite-dimensional case. As the multiverses are approximated as tensionless $p$-brane, then only possible interaction between two different tensionless p-branes is (\ref{l7}) with $a,b,c ...= 1, 2, ...N$ ($(N<D)$) labeling the different species of tensionless $p$-branes being $\mu, \nu, ...$  vectorial indices running on $1,2, ...,D$, {\it i.e.}
   \begin{equation}
     \left[ p_\mu^a, p_\nu^b\right] = i \delta_{\mu \nu} \epsilon^{ab} B, \label{l9}
   \end{equation}

   Using this interpretation and (\ref{la}) , one can compute the interaction between $N$ multiverses obtaining
   \begin{equation}
  Z=\left[ \frac{{\cal  G}_0[s(\sigma)]^{-\frac{D-3}{2}}}{\sinh\left(\frac{p\,B}{2}\int d\sigma s (\sigma)\right)}\right]^{N} ,  \label{311}
   \end{equation}
   and, therefore, each pair of multiverses  interact harmonically with frequency like $\sim B$.

   This last fact could have interesting observational consequences.  Indeed, the harmonic interaction among multiverses involves infared-ultraviolet shifts. This interaction should  produce a periodic deformation of each multiverse with period -- according to (\ref{311}) --  proportional to $1/p B$, where $p$ is the dimension of the tensionless $p$-brane.

   However, it remains to clarify the physical meaning of $B$. According to the interaction procedure sketched above, $B$ should correspond --due to dimensional reasons--  to a tiny magnetic field. The coefficient
   $\int d\sigma s (\sigma)$ is dimensionless and, therefore, one could conjecture that the only tiny magnetic field one could use as a noncommutative paremeter is the primordial magnetic field (or seed field) \cite{ru} which would be the origin of the relativistic symmetry violation.

   The numerical value for this primordial magnetic field is not presently known, but phenomenological estimations suggest that could be
   \[
   B \sim  10^{-16}\mbox{GeV}^2,
   \]
   and, therefore, the oscillation frequency of a multiverse is
   \begin{equation}
   \Omega \sim 10^{-63} \,s^{-1}.\label{eff}
   \end{equation}
   if the mass of the universe $\sim 10^{77} \,\mbox{GeV}$.

   The mechanism presented here is -in many senses- similar to the Einstein model for the heat capacity of a solid. In our case, we are taking into account only the interaction between neighbour multiverses, neglecting other interactions. Thus, the analogous to the phonon in our case is a quantum with energy $B= 10^{-8} \mbox{GeV}$.

   Thus,  an harmonic pulsation effect among multiverses could be of a new source of gravitational radiation and the relativistic invariance violation could be an explanation for the seed field puzzle \cite{ru}.

   \section{Conclusions}

   In conclusion, we have constructed the statistical mechanics of generally covariant systems moving in a noncommutative space and, from these results, we have studied the quantum statistical mechanics of tensionless membranes gas.

   We have also shown that one can  introduce no-local interactions by means of noncommutativity, which implies  measurable cosmological consequences for multiverses.

   Our main results are summarized as follows:
   \begin{itemize}
   \item{}Each null membrane is considered as a  multiverse that satisfies the cosmological principle during its evolution.  If the RHS of (\ref{5}) is zero, then the universes are causally disconnected.

   \item{}If (\ref{5}) is different from zero, then the multiverses interact harmonically implying infrared-ultraviolet shifts. The periodic fluctuations of a multiverse  could be also a source of anisotropy, maybe, they could explain the presently observed anisotropy.

   \item{}The periodic motion among multiverses is a source of gravitational waves with extremely tiny frequences.

    \item{} The noncommutativity is an effect that could be attributed to a primordial magnetic field.
   \end{itemize}

   Although the effects discussed in this paper are very small,
   in our opinion are qualitatively interesting and they can provide
   a different point of view to the standard cosmological discussions.

Finally, we would like to note that the extremely smallness value
estimated in this paper for the oscillation frequency for the
multiverses, suggests that the relativistic symmetry principle is
a good approximation for the present epoch of the universe.
However, even so one cannot discard the multiverses picture in the
early universe. 

\acknowledgments We would like to thank  J. L.
Cort\'es, A. P. Polychronakos and J. C. Retamal by discussions. This work has
been partially supported by the grants 1010596, 1010976 and
3000005 from Fondecyt-Chile and Spanish Ministerio de Educaci\'on
y Cultura grant- FPU AP07566588 and  O.N.C.E. for support.

   \end{document}